\documentclass[prb,twocolumn,showpacs,aps,superscriptaddress,floatfix]{revtex4}
\usepackage{amsmath}
\usepackage{amssymb}
\usepackage{bm}
\usepackage{graphicx}
\usepackage{color}
\usepackage{hyperref}
\usepackage{verbatim}% Long comments

\DeclareMathOperator{\sgn}{sgn}

\begin{document}

\title{Magnetic field effects in electron systems with imperfect nesting}

\author{A.O. Sboychakov}
%\affiliation{Center for Emergent Matter Science, RIKEN, Wako-shi, Saitama, 351-0198, Japan}
\affiliation{Institute for Theoretical and Applied Electrodynamics, Russian Academy of Sciences, Moscow, 125412 Russia}

\author{A.L. Rakhmanov}
%\affiliation{Center for Emergent Matter Science, RIKEN, Wako-shi, Saitama, 351-0198, Japan}
\affiliation{Institute for Theoretical and Applied Electrodynamics, Russian Academy of Sciences, Moscow, 125412 Russia}
\affiliation{Dukhov Research Institute of Automatics, Moscow, 127055 Russia}
\affiliation{Moscow Institute for Physics and Technology (State University), Moscow region, 141700 Russia}

\author{K.I. Kugel}
%\affiliation{Center for Emergent Matter Science, RIKEN, Wako-shi, Saitama, 351-0198, Japan}
\affiliation{Institute for Theoretical and Applied Electrodynamics, Russian Academy of Sciences, Moscow, 125412 Russia}
\affiliation{National Research University Higher School of Economics, Moscow, 109028 Russia}

\author{A.V. Rozhkov}
%\affiliation{Center for Emergent Matter Science, RIKEN, Wako-shi, Saitama, 351-0198, Japan}
\affiliation{Institute for Theoretical and Applied Electrodynamics, Russian Academy of Sciences, Moscow, 125412 Russia}
\affiliation{Moscow Institute for Physics and Technology (State University), Moscow region, 141700 Russia}

\author{Franco Nori}
\affiliation{Center for Emergent Matter Science, RIKEN, Wako-shi, Saitama, 351-0198, Japan}
\affiliation{Department of Physics, University of Michigan, Ann Arbor, MI 48109-1040, USA}

\begin{abstract}
We analyze the effects of an applied magnetic field on the phase diagram of a weakly-correlated electron system with imperfect nesting. The Hamiltonian under study describes two bands: electron and hole ones. Both bands have spherical Fermi surfaces, whose radii are slightly mismatched due to
doping. These types of models are often used in the analysis of magnetic states in chromium and its alloys, superconducting iron pnictides, AA-type bilayer graphene, borides, etc. At zero magnetic field, the uniform ground state of the system turns out to be unstable against electronic phase separation. The applied magnetic field affects the phase diagram in several ways. In particular, the Zeeman term stabilizes new antiferromagnetic phases. It also significantly shifts the boundaries of inhomogeneous (phase-separated) states. At sufficiently high fields, the Landau quantization gives rise to oscillations of the order parameters and of the N\'eel temperature as a function of the magnetic field.
\end{abstract}

\pacs{75.10.Lp,	
%Magnetic properties and materials: Band and itinerant models
75.50.Ee,
%Studies of specific magnetic materials: Antiferromagnetics
75.50.Cc,
%Studies of specific magnetic materials: Other ferromagnetic metals and alloys
71.20.Gj	
% Electron density of states and band structure of crystalline solids: Other metals and alloys
}

\date{\today}

\maketitle

\section{Introduction}
%%%%%%%%%%%%%%%%%%%%%%%%%%%%%%%%%%%%%%%%%%%
\label{intro}
%%%%%%%%%%%%%%%%%%%%%%%%%%%%%%%%%%%%%%%%%%%%%%%%%%

Fermi surface nesting is a very popular and important concept in
condensed matter physics~\cite{Khomskii_book2010}. The existence of two
fragments of the Fermi surface, which can be matched upon translation by a
certain reciprocal lattice vector (nesting vector), entails an instability
of a Fermi-liquid state. A superstructure or additional order parameter
related to nesting vector is generated due to the instability. The nesting
is widely invoked for the analysis of charge density wave (CDW)
states~\cite{Gruner_RMP1988_CDW,Monceau_AdvPh2012_CDW},
spin density waves (SDW)
states~\cite{Overhauser_PR1962_SDW,Gruner_RMP1994_SDW},
mechanisms of high-$T_c$
superconductivity~\cite{RuvaldsPRB1995_nesting_supercond,
GabovichSST2001_SDW-CDW_supercond,TerashimaPNAS2009_nesting_Fe_based},
fluctuating charge/orbital modulation in magnetic oxides~\cite{ChuangScience_SO_fluct},
chromium and its
alloys~\cite{shibatani_first1969,ShibataniJPSJ1969_mag_field_chromium,
shibatani1970,Rice}, etc.

It is important to emphasize that in a real material the nesting may be
imperfect, i.e. the Fermi surface fragments can only match approximately.
One of the earliest studies of imperfect nesting was performed by
Rice~\cite{Rice}
in the context of chromium and its alloys (see also the review articles
Refs.~\onlinecite{Tugushev_UFN1984_SDW_Cr,Fawcet_RMP1988_SDW_Cr}).

The notion of nesting and related concepts were broadly employed in the
recent studies of iron-based
pnictides~\cite{eremin_chub2010,Chubukov2009,graser2009,vavilov2010,
kondo2010,brydon2011,timm2012}.
For example,
Ref.~\onlinecite{eremin_chub2010}
argued that the deviation from the perfect nesting lifts degeneracy between
several competing magnetically ordered states. The influence of the
imperfect nesting on the phase coexistence was discussed in
Ref.~\onlinecite{vavilov2010}.

Many theoretical investigations assume from the outset the homogeneity of
the electron state. This assumption may be violated in systems with
imperfect nesting. Indeed, it was demonstrated that the imperfect-nesting
mechanism can be responsible for the nanoscale phase
separation in quasi-one-dimensional
metals~\cite{tokatly1992},
chromium
alloys~\cite{WeImperf},
iron-based
superconductors~\cite{Sboychakov_PRB2013_PS_pnict},
and in doped bilayer
graphene~\cite{ourBLGreview, Sboychakov_PRB2013_MIT_AAgraph,
Sboychakov_PRB2013_PS_AAgraph}.
Several experiments on
pnictides~\cite{PSexp1,PSexp2,PSexp3,goko2009,phasep_exp2012,phasep_exp2014,
phasep_exp2016}
and
chalcogenides~\cite{PSexp4,bianconi_phasep2011,bianconi_phasep2015}
support the possibility of phase separation (see also review
article~[\onlinecite{Dagotto_review2012}]).

In similar context of imperfect nesting, studies of spin and charge
inhomogeneities are currently active in
the physics of low-dimensional
compounds.~\cite{Narayanan_RRL2014,Campi_Nature2015,Chen_PRB2014}
Other types of inhomogeneous states (``stripes'', domain walls, impurity
levels) were also discussed in the literature in the framework of analogous
models~\cite{zaanen_stripes1989,tokatly1992,Akzyanov2015}.
Moreover, it was shown that the possibility of SDW ordering in systems with
itinerant charge carriers results in very rich and complicated phase
diagrams involving phase-separation
regions~\cite{Igoshev_JPCM2015,Igoshev_JMMM2015}.

An applied magnetic field $\mathbf{B}$ alters the quasiparticle states, changing the nesting conditions. In the present paper, we explore the physical consequences of the applied magnetic field for weakly-correlated electron systems with imperfect nesting. In a generic situation, the magnetic field enters the Hamiltonian both via the Zeeman term, and via the substitution
$\hat{\mathbf{p}} \rightarrow \hat{\mathbf{p}}+(e/c)\mathbf{A}$.
The Zeeman term lifts the degeneracy with respect to the spin projection. Both electron and hole Fermi surface sheets become split into two spin-polarized components. As a result, two different SDW order parameters corresponding to spin projections parallel and antiparallel to the direction of $\mathbf{B}$ can be constructed. When the electron-hole symmetry between the electron and hole pockets is absent, the effects of the Zeeman term is especially pronounced. In particular, new antiferromagnetic (AFM) phases, both homogeneous and inhomogeneous, appear in the phase diagram. The boundaries between different phases exhibit strong dependence on $B$. For example, such effects were found in the analysis of the magnetic phase diagram of doped rare-earth borides~\cite{Sluchanko_PRB2015}.

We also study separately phenomena caused by the Landau quantization. In the range of high magnetic fields, Landau quantization leads to
characteristic oscillations of the SDW order parameters and of the N\'eel temperature as a function of $B$. However, these oscillations are most clearly pronounced in the case of symmetric electron and hole pockets. Otherwise, the nesting and, hence, the SDW ordering would be completely destroyed by the magnetic field before any detectable oscillations would occur. Similar oscillatory effects are well-known in the context of quasi-one-dimensional compounds~\cite{GorkovLebed_JdePhL1984_magn_field_stab1D,
Lebed_JETP1985_magn_field_PhDiag,Yakovenko_JETP1987_field_ind_PhTr,
Azbel_PRB1989_field_ind_SDW,Pudalov_Springer2008_magn_field_SDW,
Lebed_Springer2008_magn_field_SDW}. However, the role of the magnetic field in  nesting-related phenomena in the usual three-dimensional materials has received only limited attention~\cite{YamasakiJPSJ1967_mag_field_SDW,
ShibataniJPSJ1969_mag_field_chromium}.

This paper is organized as follows. In Section~\ref{model}, we formulate the model. Section~\ref{SectLowB} deals with the effects related to the Zeeman term. Phenomena occurring due to the Landau quantization are treated in Section~\ref{SectHighB}. A discussion of the results is given in Section~\ref{disco}. Some details of the calculations are presented in the Appendix.

\section{Model}
%%%%%%%%%%%%%%%%%%%%%%%%%%%%%%%%%%%%%%%%%%%
\label{model}
%%%%%%%%%%%%%%%%%%%%%%%%%%%%%%%%%%%%%%%%%%%%%%%%%%

\subsection{Hamiltonian}
\begin{figure}[t] \centering
\includegraphics[width=0.9\columnwidth]{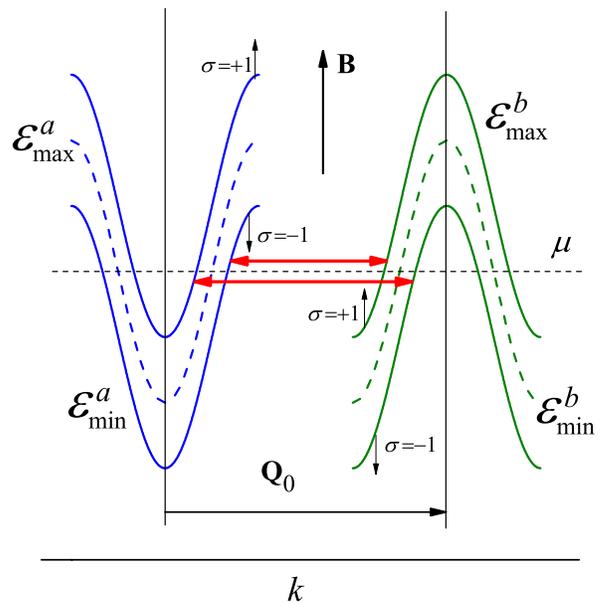}
\caption{(Color online) Band structure of the electron model in an applied
magnetic field. The magnetic field lifts the degeneracy of the
electron-like ($a$) and hole-like ($b$) bands with respect to the electron
spin. The red arrows indicate the interband coupling giving rise to the
order parameters. The splitting into Landau levels is not shown.
%%%%%%%%%%%%%%%%%%%%%%%%%%%%%%%%%%%%%%%%%%%
\label{band}
%%%%%%%%%%%%%%%%%%%%%%%%%%%%%%%%%%%%%%%%%%%%%%%%%%
}
\end{figure}

%%%%%%%%%%%%%%%%%%%%%%%%%%%%%%%%%%%%%%%%%%%%%%%%%%
\label{subsect::hamilt}
%%%%%%%%%%%%%%%%%%%%%%%%%%%%%%%%%%%%%%%%%%%%%%%%%%
The model under study is schematically illustrated in Fig.~\ref{band}.
It describes two bands: an electronic band ($a$) and a hole band ($b$). The
hole Fermi surface coincides with the electron Fermi surface after a
translation by a reciprocal lattice vector $\mathbf{Q}_0$.
The quasiparticles interact with each other via a short-range repulsive
potential. Formally, the Hamiltonian is represented as
\begin{equation}
%%%%%%%%%%%%%%%%%%%%%%%%%%%%%%%%%%%%%%%%%%%
\label{ham_summa}
%%%%%%%%%%%%%%%%%%%%%%%%%%%%%%%%%%%%%%%%%%%%%%%%%%
\hat{H}=\hat{H}_e+\hat{H}_{\textrm{int}}\,,
\end{equation}
where $\hat{H}_e$ is the single-electron term, and $\hat{H}_{\textrm{int}}$
corresponds to the interaction between quasiparticles.

Regarding the single-electron term, we assume a quadratic dispersion for both
bands and use the Wigner--Seitz approximation. Specifically, in the
electron band, the wave vector $\mathbf{k}$ is confined within a sphere of
finite radius centered at zero, and in the hole band, such sphere is
centered at
$\mathbf{Q}_0$.
The kinetic energies of these states are spread between the minimum values (denoted by $\varepsilon^{a,b}_{\textrm{min}}$)
and maximum values $\varepsilon^{a,b}_{\textrm{max}}$,
see Fig.~\ref{band}. Thus, the energy spectra for the electron and hole pockets, measured relative to the Fermi energy $\mu$, have the form ($\hbar = 1$)
\begin{eqnarray}
%%%%%%%%%%%%%%%%%%%%%%%%%%%%%%%%%%%%%%%%%%%
\label{spectra}
%%%%%%%%%%%%%%%%%%%%%%%%%%%%%%%%%%%%%%%%%%%%%%%%%%
&&\varepsilon^a(\mathbf{k})
=
\frac{\mathbf{k}^2}{2m_a} + \varepsilon^{a}_{\textrm{min}}-\mu,
\quad
\varepsilon^{a}_{\textrm{min}}<\varepsilon^a<\varepsilon^{a}_{\textrm{max}},
\\
&&\varepsilon^b(\mathbf{k}+\mathbf{Q}_0)
=
-\frac{\mathbf{k}^2}{2m_b}+\varepsilon^{b}_{\textrm{max}}-\mu,
\quad
\varepsilon^{b}_{\textrm{min}} < \varepsilon^b
<
\varepsilon^{b}_{\textrm{max}}.
\nonumber
\end{eqnarray}
The nesting conditions mean that for some $\mu=\mu_0$,
the Fermi surfaces of the $a$ and $b$ bands coincide after a translation by
the vector $\mathbf{Q}_0$, and both Fermi spheres are characterized by the single Fermi momentum $k_F$. Using Eqs.~\eqref{spectra}, we readily obtain
\begin{equation}
%%%%%%%%%%%%%%%%%%%%%%%%%%%%%%%%%%%%%%%%%%%
\label{nesting}
%%%%%%%%%%%%%%%%%%%%%%%%%%%%%%%%%%%%%%%%%%%%%%%%%%
\!\!k_F^2\!=\!\frac{2m_am_b}{m_a\!+\!m_b}\left(\varepsilon^{b}_{\textrm{max}}
\!-\!\varepsilon^{a}_{\textrm{min}}\right),\,\, \mu_0\!=\!\frac{m_b\varepsilon^{b}_{\textrm{max}}\!+\!m_a
\varepsilon^{a}_{\textrm{min}}}{m_a\!+\!m_b}.
\end{equation}
Below, we will measure the momentum of the $b$ band from the nesting vector
$\mathbf{Q}_0$, that is, we replace $\varepsilon^b(\mathbf{k}+\mathbf{Q}_0)\to\varepsilon^b(\mathbf{k})$
in Eq.~\eqref{spectra}.
For perfect electron--hole symmetry, when $m_a=m_b=m$ and $\varepsilon^{b}_{\textrm{max}}=-\varepsilon^{a}_{\textrm{min}}$,
we obtain $\mu_0=0$.

In an applied uniform dc magnetic field $\mathbf{B}$, the single-electron
part of our model can be written as
\begin{equation}
%%%%%%%%%%%%%%%%%%%%%%%%%%%%%%%%%%%%%%%%%%%
\label{ham_e}
%%%%%%%%%%%%%%%%%%%%%%%%%%%%%%%%%%%%%%%%%%%%%%%%%%
\hat{H}_e=\sum_{\alpha\sigma}\int\!d^3x\,\psi^\dag_{\alpha\sigma}
(\mathbf{x})\hat{H}_{\alpha \sigma}\psi_{\alpha\sigma}(\mathbf{x})\,,
\end{equation}
where [see Eqs.~\eqref{spectra}]
\begin{eqnarray}
%%%%%%%%%%%%%%%%%%%%%%%%%%%%%%%%%%%%%%%%%%%
\label{ham_e_ham}
%%%%%%%%%%%%%%%%%%%%%%%%%%%%%%%%%%%%%%%%%%%%%%%%%%
\nonumber
\hat{H}_{a\sigma}
&=&
\frac{\left(\hat{\mathbf{p}}+\frac{e}{c}\mathbf{A}\right)^2}{2m_a}
+
\sigma g_a\omega_a
+
\varepsilon_{\textrm{min}}^a-\mu,
\\
\hat{H}_{b\sigma}
&=&
-\frac{\left(\hat{\mathbf{p}}+\frac{e}{c}\mathbf{A}\right)^2}{2m_b}
+
\sigma g_b\omega_b+\varepsilon_{\textrm{max}}^b-\mu.
\end{eqnarray}
In these equations,
$\alpha=a,b$,
$\mathbf{\hat{p}}=-i\bm{\nabla}$
is the momentum operator,
$\sigma=\pm 1$
is the spin projection, $\omega_\alpha=eB/cm_\alpha$
are the cyclotron frequencies for the electron and hole bands, and
$g_{\alpha}$ are the corresponding Land\'e factors. We assume that the magnetic field is directed along the $z$ axis and choose the Landau gauge for the vector potential, $\mathbf{A}=(-By,0,0)$.

The second term in Eq.~\eqref{ham_summa} describes the interaction between electrons and holes. For treating the SDW instability, it is sufficient to keep only the interaction between the $a$ and $b$ bands. The neglected intraband contributions can only renormalize the parameters. We also assume that this interaction is a short-range one. Thus, we can write
\begin{equation}
%%%%%%%%%%%%%%%%%%%%%%%%%%%%%%%%%%%%%%%%%%%
\label{ham_int}
%%%%%%%%%%%%%%%%%%%%%%%%%%%%%%%%%%%%%%%%%%%%%%%%%%
\hat{H}_{\textrm{int}}\!
=\!
V\sum_{\sigma\sigma'}
\int\!{d^3x\,
	\psi^\dag_{a\sigma}(\mathbf{x})
	\psi_{a\sigma}(\mathbf{x})
	\psi^\dag_{b\sigma'}(\mathbf{x})
	\psi_{b\sigma'}(\mathbf{x})
      }\,.
\end{equation}
The coupling constant $V$ is positive, which corresponds to repulsion.

\subsection{Single-electron spectrum in a magnetic field}
%%%%%%%%%%%%%%%%%%%%%%%%%%%%%%%%%%%%%%%%%%%
\label{subsect::non-interact}
%%%%%%%%%%%%%%%%%%%%%%%%%%%%%%%%%%%%%%%%%%%%%%%%%%

Let us start with a brief discussion of the properties of the single-electron Hamiltonian. When the magnetic field is zero, the single-electron spectrum consists of two bands of free fermions with two-fold spin degeneracy. In a non-zero applied magnetic field ${\bf B}$, the operator $\psi_{\alpha\sigma}(\mathbf{x})$ can be expressed as a series expansion in terms of eigenfunctions of Hamiltonian \eqref{ham_e},
\begin{equation}
%%%%%%%%%%%%%%%%%%%%%%%%%%%%%%%%%%%%%%%%%%%
\label{psi_expand}
%%%%%%%%%%%%%%%%%%%%%%%%%%%%%%%%%%%%%%%%%%%%%%%%%%
\psi_{\alpha\sigma}(\mathbf{x})
=
\sum_{\mathbf{p}n}
	{\frac{e^{i(p_xx+p_zz)}}{\sqrt{{\cal V}^{2/3}l_B}}}\,
	\chi_n\!\! \left(\frac{y-p_xl_B^2}{l_B}\right)
	\psi_{\mathbf{p}n\alpha\sigma}\,,
\end{equation}
where
$\psi^{\phantom{\dag}}_{\mathbf{p}n\alpha\sigma}$
is the annihilation operator for an electron in band $\alpha$ with 2D
momentum 
$\mathbf{p}=(p_x,p_z)$
and spin projection $\sigma$ at the Landau level $n$, symbol
${\cal V}$
denotes the system volume,
$l_B=\sqrt{c/eB}$
is the magnetic length,
\begin{equation}
%%%%%%%%%%%%%%%%%%%%%%%%%%%%%%%%%%%%%%%%%%%
\label{hi_n}
%%%%%%%%%%%%%%%%%%%%%%%%%%%%%%%%%%%%%%%%%%%%%%%%%%
\chi_n(\xi)=\frac{1}{\sqrt{2^nn!\sqrt{\pi}}}e^{-\xi^2/2}H_n(\xi)\,,
\end{equation}
and $H_n(\xi)$ is the Hermite polynomial of degree $n$. In this basis, the Hamiltonian can be expressed as
\begin{equation}
%%%%%%%%%%%%%%%%%%%%%%%%%%%%%%%%%%%%%%%%%%%
\label{ham_e1}
%%%%%%%%%%%%%%%%%%%%%%%%%%%%%%%%%%%%%%%%%%%%%%%%%%
\hat{H}_e
=
\sum_{\mathbf{p}n\alpha\sigma}\!
	\varepsilon_{\alpha \sigma}(p_z,n)
	\psi^\dag_{\mathbf{p}n\alpha\sigma}
	\psi^{\phantom{\dag}}_{\mathbf{p}n\alpha\sigma}\,,
\end{equation}
where the single-particle eigenenergies are
\begin{eqnarray}
%%%%%%%%%%%%%%%%%%%%%%%%%%%%%%%%%%%%%%%%%%%
\label{E_alpha}
%%%%%%%%%%%%%%%%%%%%%%%%%%%%%%%%%%%%%%%%%%%%%%%%%%
\varepsilon^a_{\sigma}(p_z,n)
&\!\!=\!&\!
\omega_a\!\left(n+\frac{1}{2}+\sigma g_a\right)
+
\frac{p_z^2}{2m_a}+\varepsilon^{a}_{\textrm{min}}-\mu,
\\
\varepsilon^b_{\sigma}(p_z,n)
&\!\!=\!&\!
-\omega_b\!\left(n+\frac{1}{2}-\sigma g_b\right)
-\frac{p_z^2}{2m_b}+\varepsilon^{b}_{\textrm{max}}-\mu.
\nonumber
\end{eqnarray}
The spectrum consists of four bands (see Fig.~\ref{band}) since the Zeeman term (the term, proportional to $\sigma$) lifts the degeneracy with respect to the electron spin.

\subsection{Energy scales}
%%%%%%%%%%%%%%%%%%%%%%%%%%%%%%%%%%%%%%%%%%%%%%%%%%
\label{subsect::scales}
%%%%%%%%%%%%%%%%%%%%%%%%%%%%%%%%%%%%%%%%%%%%%%%%%%

The energy spectrum of the model is characterized by two single-particle
energy scales. The first is the Fermi energy
$\varepsilon_{F\alpha}=k_F^2/2m_\alpha$, and the second is
$\omega_{\alpha}$, which is the distance between the Landau levels in band
$\alpha$. Furthermore, we assume that $\varepsilon_{Fa}\approx
\varepsilon_{Fb}$ and $\omega_{a}\approx\omega_{b}$. The energy scale
associated with the interactions will be characterized by the value of a
spectral gap $\Delta_0$. The latter parameter is defined as follows. When
$\mu\approx\mu_0$, the nesting between the two sheets of the Fermi surface
is nearly perfect. It is known that, under such condition, the interaction
between the electron- and hole-like bands opens a gap $\Delta(T,B)$ in the
electron spectrum. The value of the gap at zero temperature $T=0$ and zero
magnetic field $B=0$ will be denoted as
$\Delta_0 = \Delta(0,0)$.

Below we consider the case $\Delta_0\ll\varepsilon_{F\alpha}$,
which corresponds to a weak electron-hole coupling. We also classify the
magnetic field as low if $\omega_{\alpha}\lesssim\Delta_0$, and high if
$\omega_{\alpha}\gtrsim\Delta_0$. The Landau quantization is of importance in the high-field range, whereas at low fields, it can be neglected. In the regime of low magnetic fields considered in the next Section~\ref{SectLowB},
we neglect any corrections associated with the small ratio $\omega_{\alpha}/\varepsilon_{F\alpha}$, while for $\omega_{\alpha}\gtrsim\Delta_0$ (this regime is considered in
Section~\ref{SectHighB}), we take into account these corrections in the leading order, which turns out to be of the order of $(\omega_{\alpha}/\varepsilon_{F\alpha})^{1/2}$.

\section{Electron-hole coupling: Low magnetic field}
%%%%%%%%%%%%%%%%%%%%%%%%%%%%%%%%%%%%%%%%%%%
\label{SectLowB}
%%%%%%%%%%%%%%%%%%%%%%%%%%%%%%%%%%%%%%%%%%%%%%%%%%

\subsection{Main definitions}

At low magnetic fields, we can neglect the effect of the Landau
quantization on the electron spectrum and take into account only the Zeeman
splitting. In this approximation, the single-electron Hamiltonian \eqref{ham_e1} has the form
\begin{equation}
%%%%%%%%%%%%%%%%%%%%%%%%%%%%%%%%%%%%%%%%%%%
\label{SEHam}
%%%%%%%%%%%%%%%%%%%%%%%%%%%%%%%%%%%%%%%%%%%%%%%%%%
\hat{H}_e=\sum_{\mathbf{k}\alpha\sigma}\!\varepsilon_{\alpha\sigma}
(\mathbf{k})\psi^\dag_{\mathbf{k}\alpha\sigma}
\psi^{\phantom{\dag}}_{\mathbf{k}\alpha\sigma}\,,
\end{equation}
where $\psi^\dag_{\mathbf{k}\alpha\sigma}$ and
$\psi^{\phantom{\dag}}_{\mathbf{k}\alpha\sigma}$ are the creation and annihilation operators of an electron in band $\alpha$ with (3D) momentum
$\mathbf{k}$ and spin projection $\sigma$, while the electron spectra now read
%\begin{eqnarray}
%%%%%%%%%%%%%%%%%%%%%%%%%%%%%%%%%%%%%%%%%%%%
%\label{E_alpha-LowB}
%%%%%%%%%%%%%%%%%%%%%%%%%%%%%%%%%%%%%%%%%%%%%%%%%%%
%\varepsilon^a_{\sigma}(\mathbf{k})&=& \frac{\mathbf{k}^2}{2m_a}+ \sigma g_a\omega_a +\varepsilon^{a}_{\textrm{min}}-\mu\,,\\
%\varepsilon^b_{\sigma}(\mathbf{k})&=&-\frac{\mathbf{k}^2}{2m_b}+ \sigma g_b\omega_b +\varepsilon^{b}_{\textrm{max}}-\mu\,.\nonumber
%\end{eqnarray}
%
%Further on, we are interested in the ordered state of our system, which
%exists only in the parameter range close to perfect nesting. So, we rewrite
%Eqs.~\eqref{E_alpha-LowB}
%in terms of
%$k_F$
%and
%$\mu_0$,
%Eq.~\eqref{nesting},
\begin{eqnarray}
%%%%%%%%%%%%%%%%%%%%%%%%%%%%%%%%%%%%%%%%%%%
\label{E_alpha-LowB_kF}
%%%%%%%%%%%%%%%%%%%%%%%%%%%%%%%%%%%%%%%%%%%%%%%%%%
\nonumber
\varepsilon^a_{\sigma}(\mathbf{k})
&=&
\frac{k^2-k^2_F}{2m_a}+ \sigma g_a\omega_a-\delta\mu\,,
\\
\varepsilon^b_{\sigma}(\mathbf{k})
&=&
-\frac{k^2-k^2_F}{2m_b}+ \sigma g_b\omega_b-\delta\mu\,,
\end{eqnarray}
where
$\delta\mu=\mu-\mu_0$.

If the applied magnetic field is zero, the commensurate SDW %antiferromagnetic (AFM)
order parameter can be written as
\begin{equation}
%%%%%%%%%%%%%%%%%%%%%%%%%%%%%%%%%%%%%%%%%%%
\label{rice}
%%%%%%%%%%%%%%%%%%%%%%%%%%%%%%%%%%%%%%%%%%%%%%%%%%
\Delta=\frac{V}{{\cal V}}
\sum_{\mathbf{k}}
	\left\langle
		\psi^\dag_{\mathbf{k}a\sigma}
		\psi^{\phantom{\dag}}_{\mathbf{k}b\bar{\sigma}}
	\right\rangle\,,
\end{equation}
where $\bar{\sigma}$ means $-\sigma$. This order parameter is degenerate with respect to spin. If ${\bf B} \ne 0$, this degeneracy is lifted and we introduce a two-component order parameter corresponding to the nesting vectors shown by the arrows in Fig.~\ref{band}
\begin{equation}
%%%%%%%%%%%%%%%%%%%%%%%%%%%%%%%%%%%%%%%%%%%
\label{order_lowB}
%%%%%%%%%%%%%%%%%%%%%%%%%%%%%%%%%%%%%%%%%%%%%%%%%%
\Delta_\uparrow
\!=\!
\frac{V}{{\cal V}}\!
\sum_{\mathbf{k}}\!
	\left\langle\!
		\psi^\dag_{\mathbf{k}a\uparrow}
		\psi^{\phantom{\dag}}_{\mathbf{k}b\downarrow}\!
	\right\rangle,\,\,\,
\Delta_\downarrow
\!=\!
\frac{V}{{\cal V}}\!
\sum_{\mathbf{k}}\!
	\left\langle\!
		\psi^\dag_{\mathbf{k}a\downarrow}
		\psi^{\phantom{\dag}}_{\mathbf{k}b\uparrow}\!
	\right\rangle\,.
\end{equation}
The mean-field spectrum of the model has a form
\begin{equation}
%%%%%%%%%%%%%%%%%%%%%%%%%%%%%%%%%%%%%%%%%%%
\label{Emf_LB}
%%%%%%%%%%%%%%%%%%%%%%%%%%%%%%%%%%%%%%%%%%%%%%%%%%
E^\sigma_{1,2}(\mathbf{k})
=
\frac{\varepsilon_\sigma^a(\mathbf{k})
	+
	\varepsilon_{-\sigma}^b(\mathbf{k})}{2}
\pm
\sqrt{\Delta_\sigma^2
	+\!
	\left(
		\frac{\varepsilon_\sigma^a(\mathbf{k})
			-
			\varepsilon_{-\sigma}^b(\mathbf{k})
		}{2}
	\right)^{\!\!2}
}.
\end{equation}
Using these spectra, we can write the grand potential of the system in the
mean-field approximation as a sum of two ``decoupled'' terms
$\Omega = \Omega_\uparrow + \Omega_\downarrow$,
where ``partial'' grand potentials are equal to
\begin{eqnarray}
%%%%%%%%%%%%%%%%%%%%%%%%%%%%%%%%%%%%%%%%%%%
\label{GrPot}
%%%%%%%%%%%%%%%%%%%%%%%%%%%%%%%%%%%%%%%%%%%%%%%%%%
\Omega_\sigma
=
{\cal V}\left[\!\frac{\Delta^2_\sigma}{V}-
T\!\!\sum_{s=1,2}\int\!\frac{d^3\mathbf{k}}{(2\pi)^3}
\ln{\!\left(1\!+\!e^{-E_s^\sigma(\mathbf{k})/T}\right)}\!\right]\!.
\end{eqnarray}
The order parameters are found by minimizing $\Omega$ with respect to $\Delta_{\sigma}$.

\subsection{SDW order parameters}

The case when the electron and hole bands are perfectly symmetric is, of
course, the simplest. In such a situation, however, the effect of weak
magnetic fields on the electron spectrum is zero, as it will be evident below. Thus, we should introduce some electron-hole asymmetry to obtain non-trivial results in the low-field range. Qualitatively, a particular source of the asymmetry is not of importance. Here, we assume for simplicity that $m_a=m_b=m$ (hence, $\omega_a=\omega_b=\omega_H$ and
$\varepsilon_{Fa}=\varepsilon_{Fb}=\varepsilon_F$), but $g_a\neq g_b$. It is also assumed that the difference $g_a-g_b$ is of the same order as $g_a$
and $g_b$.

We rewrite Eqs.~\eqref{E_alpha-LowB_kF} in the following convenient form
\begin{eqnarray}
%%%%%%%%%%%%%%%%%%%%%%%%%%%%%%%%%%%%%%%%%%%
\label{E_alpha-LowB_ksig}
\varepsilon^a_{\sigma}(\mathbf{k})
&=&
\left(\frac{k^2}{2m}-E_{F\sigma}\right)-\mu_\sigma,
\nonumber
\\
\varepsilon^b_{-\sigma}(\mathbf{k})
&=&
-\left(\frac{k^2}{2m} -E_{F\sigma}\right)-\mu_\sigma,
\end{eqnarray}
where the following notation is used
\begin{eqnarray}
%%%%%%%%%%%%%%%%%%%%%%%%%%%%%%%%%%%%%%%%%%%
\label{notations}
g &=& \frac{g_a+g_b}{2},\quad \Delta g = \frac{g_a-g_b}{2},
\nonumber
\\
E_{F\sigma}
&=&
\frac{k_F^2}{2m}-\sigma g\omega_H,
\quad
\mu_\sigma=\delta\mu-\sigma \Delta g\omega_H.
\end{eqnarray}
As we stated above, in the low-field range, we neglect corrections of the
order of $\omega_H/E_F$, since $\omega_H\ll\Delta_0\ll E_F$. Then, we take into account only terms of the order of  $\omega_H/\Delta_0$. Expanding the spectra in Eqs.~\eqref{E_alpha-LowB_ksig} near the Fermi momentum, we obtain
\begin{eqnarray}
%%%%%%%%%%%%%%%%%%%%%%%%%%%%%%%%%%%%%%%%%%%
\label{pmE_PEH_LB}
%%%%%%%%%%%%%%%%%%%%%%%%%%%%%%%%%%%%%%%%%%%%%%%%%%
\varepsilon_\sigma^a(\mathbf{k})
&\approx&
v_F\delta k\, + \sigma g \omega_H -\mu_\sigma\,,
\\
\nonumber
\varepsilon_{-\sigma}^b(\mathbf{k})
&\approx&
-v_F\delta k\, - \sigma g \omega_H -\mu_\sigma\,,
\end{eqnarray}
where
$\delta k=|\mathbf{k}|-k_F$
and
$v_F=k_F/m$.

Substituting Eqs.~\eqref{pmE_PEH_LB} in Eqs.~\eqref{Emf_LB} and \eqref{GrPot} and performing integration, we obtain the expression for grand potential
\begin{eqnarray}
%%%%%%%%%%%%%%%%%%%%%%%%%%%%%%%%%%%%%%%%%%%
\label{GrPot_LB_EHS}
&&\frac{\Omega}{{\cal V}}=2N_F
\sum_{\sigma}\left[-\frac{\Delta^2_\sigma}{2}\left(\ln{\frac{\Delta_0}
{\Delta_\sigma}}+\frac{1}{2}\right)+\phantom{\int\limits_0^\infty}\right.\\
&&\left.T\!\int\limits_0^\infty\!\! d\xi\ln{\left[f_F(\sqrt{\Delta_{\sigma}^2+\xi^2}-\mu_\sigma)f_F
(\sqrt{\Delta_{\sigma}^2+\xi^2}+\mu_\sigma)\right]}\right]\!,\nonumber
\end{eqnarray}
where
$f_F(\epsilon)=1/[1+\exp{(\epsilon/T)}]$
is the Fermi function, and $\Delta_0$ is the SDW gap at zero field, temperature, and doping ($\mu=\mu_0$)
\begin{equation}
%%%%%%%%%%%%%%%%%%%%%%%%%%%%%%%%%%%%%%%%%%%
\label{BCS}
\Delta_0\approx\varepsilon_F\exp{(-1/VN_F)},\quad N_F=\frac{k_F^2}{2\pi^2v_F}.
\end{equation}
From the minimization conditions $\partial\Omega/\partial\Delta_{\sigma}=0$,
we derive equations for the order parameters
\begin{equation}
%%%%%%%%%%%%%%%%%%%%%%%%%%%%%%%%%%%%%%%%%%%
\label{Delta_LF-EHS}
\!\!\ln{\frac{\Delta_0}{\Delta_\sigma}}\!=\!\!\int\limits_0^\infty\!\!d\xi
\frac{f_F(\!\sqrt{\Delta_{\sigma}^2\!+\!\xi^2}\!+\!\mu_\sigma\!)\!+
\!f_F(\!\sqrt{\Delta_{\sigma}^2\!+\!\xi^2}\!-\!\mu_\sigma\!)}
{\sqrt{\Delta_{\sigma}^2+\xi^2}}.
\end{equation}

The electron density is
\begin{equation}
%%%%%%%%%%%%%%%%%%%%%%%%%%%%%%%%%%%%%%%%%%%
\label{n}
N=\frac{1}{{\cal V}}\sum_{\mathbf{k}s\sigma}{f_F[E^\sigma_s(\mathbf{k})]}\,.
\end{equation}
The parameter $N_0$ corresponds to the ideal nesting, $\delta\mu=0$, in the absence of the magnetic field, $B=0$. We define the doping level as $X=N-N_0$.
The equation for $X$ can be written in the form
\begin{equation}
%%%%%%%%%%%%%%%%%%%%%%%%%%%%%%%%%%%%%%%%%%%
\label{n_LF_EHS}
\frac{X}{N_F}\!=\!\!\!\sum_\sigma\!\!\int\limits_0^\infty\!\!d\xi\!
\left[\!f_F(\!\sqrt{\Delta_{\sigma}^2\!+\!\xi^2}\!-\!\mu_\sigma\!)\!
-\!\!f_F(\!\sqrt{\Delta_{\sigma}^2\!+\!\xi^2}\!+\!\mu_\sigma\!)\!\right].
\end{equation}
The derivation of this equation is straightforward (the details can be found in Ref.~\onlinecite{WeImperf}). The value of $X$ can also be considered as a shift from the position of ideal nesting. In our terms, ``zero doping'' really means ``perfect nesting''.

For further calculations, it is convenient to introduce the following
dimensionless variables
\begin{equation}
%%%%%%%%%%%%%%%%%%%%%%%%%%%%%%%%%%%%%%%%%%%
\label{dimen}
x\!=\!\frac{X}{2N_F\Delta_0},\,\,
\nu\!=\!\frac{\delta\mu}{\Delta_0},\,\,
b\!=\!\frac{\Delta g\omega_H}{\Delta_0},\,\,
\delta_\sigma\!=\!\frac{\Delta_\sigma}{\Delta_0}\,.
\end{equation}
In this notation, we rewrite Eqs.~\eqref{Delta_LF-EHS} and~\eqref{n_LF_EHS} as
\begin{eqnarray}
%%%%%%%%%%%%%%%%%%%%%%%%%%%%%%%%%%%%%%%%%%%
\label{dimenEq_LF_EHS}
\ln{\frac{1}{\delta_\sigma}}\!\!&=&\!\!\int\limits_0^\infty\!\!
\frac{d\xi}{\eta_{\sigma}}\left[f_F(\eta_{\sigma}\!+\!\nu\!-\!\sigma b)\!+\!f_F(\eta_{\sigma}\!-\!\nu\!+\!\sigma b)\right],\\
\!\!x\!&=&\!\!\int\limits_0^\infty\!\!d\xi\sum_\sigma
\left[f_F(\eta_{\sigma}\!-\!\nu\!+\!\sigma b)\!-\!f_F(\eta_{\sigma}\!+\!\nu\!-\!\sigma b)\right],\nonumber
\end{eqnarray}
where
$\eta_{\sigma}=\sqrt{\delta_\sigma^2+\xi^2}$.
We also introduce the dimensionless grand potential
\begin{equation}
%%%%%%%%%%%%%%%%%%%%%%%%%%%%%%%%%%%%%%%%%%%
\label{GP_dim}
\varphi=\frac{\pi^2v_F}{k_F^2\Delta_0^2}\frac{\Omega}{{\cal V}}\,.
\end{equation}
Using notation Eqs.~\eqref{dimen}, we rewrite Eq.~\eqref{GrPot_LB_EHS} in the dimensionless form
\begin{eqnarray}
%%%%%%%%%%%%%%%%%%%%%%%%%%%%%%%%%%%%%%%%%%%
\label{GrPot_LB_EHS_dim}
%%%%%%%%%%%%%%%%%%%%%%%%%%%%%%%%%%%%%%%%%%%%%%%%%%
\varphi
&=&
\sum_\sigma\varphi_\sigma
=\!
\sum_\sigma
	\left\{
		-\frac{\delta_\sigma^2}{2}
		\left(\ln{\frac{1}{\delta_\sigma}}+\frac{1}{2}\right)
		+
		\phantom{\int\limits_0^\infty}
		\right.\\
	  	&& \left.
		t\int\limits_0^\infty\!\! d\xi
			\ln{\left[
				f_F(\eta_\sigma+\nu-\sigma b)
				f_F(\eta_\sigma-\nu+\sigma b)
			\right]}
	\right\}\,,
\nonumber
\end{eqnarray}
where
$t=T/\Delta_0$.

We need to consider the system at fixed doping rather than at fixed chemical potential. Such a choice is better suited for describing usual experimental conditions. To work at fixed $x$, we should calculate the system's free energy $f=\varphi+\nu x$. To do this, we solve the system of
equations~\eqref{dimenEq_LF_EHS} at a given doping level $x$. Then, we calculate $\varphi$ and $f$ using the obtained values of $\delta_\sigma$ and $\nu$.

In the paramagnetic state, $\delta_\uparrow=\delta_\downarrow=0$, we readily find from Eqs.~\eqref{dimenEq_LF_EHS} and~\eqref{GrPot_LB_EHS_dim} that the chemical potential is proportional to the doping $\nu=x/2$, and
\begin{eqnarray}
%%%%%%%%%%%%%%%%%%%%%%%%%%%%%%%%%%%%%%%%%%%
\label{OmegaFPM}
%%%%%%%%%%%%%%%%%%%%%%%%%%%%%%%%%%%%%%%%%%%%%%%%%%
\varphi&=&-\nu^2-b^2-\frac{\pi^2t^2}{3}\,,\nonumber\\
f&=&\frac{x^2}{4}-b^2-\frac{\pi^2t^2}{3}\,.
\end{eqnarray}
The properties of the ordered phases will be discussed below.

\subsection{Homogeneous phases at zero temperature}
%%%%%%%%%%%%%%%%%%%%%%%%%%%%%%%%%%%%%%%%%%%%%%%%%%
\label{subsect::homogen}
%%%%%%%%%%%%%%%%%%%%%%%%%%%%%%%%%%%%%%%%%%%%%%%%%%
First, let us discuss the homogeneous phases allowed by our mean-field scheme. In what follows, we will limit ourselves to the case $T=0$.

The task is simplified by the fact that in the mean-field approach, our system becomes ``decoupled'' and consists of two independent subsystems, labeled by the index $\sigma$. The order parameters of these subsystems are mutually independent [see Fig.~\ref{band}, Eqs.~(\ref{order_lowB}) and~(\ref{GrPot})]. For such a situation, the thermodynamic phases of the system are characterized by a pair of order parameters $(\Delta_\uparrow, \Delta_\downarrow)$.

\begin{figure}[t]
\centering
\includegraphics[width=0.95\columnwidth]{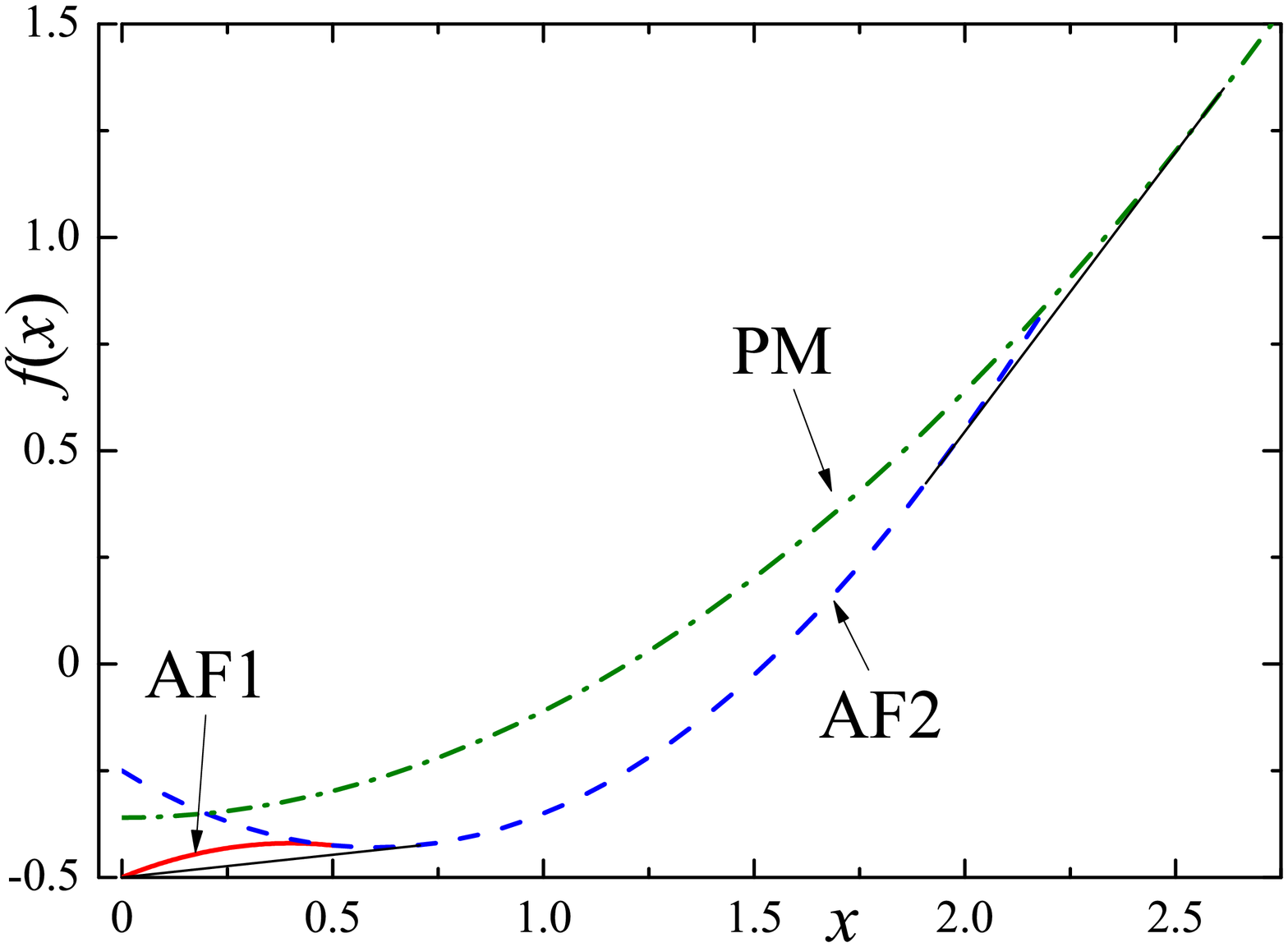}\\
\includegraphics[width=0.95\columnwidth]{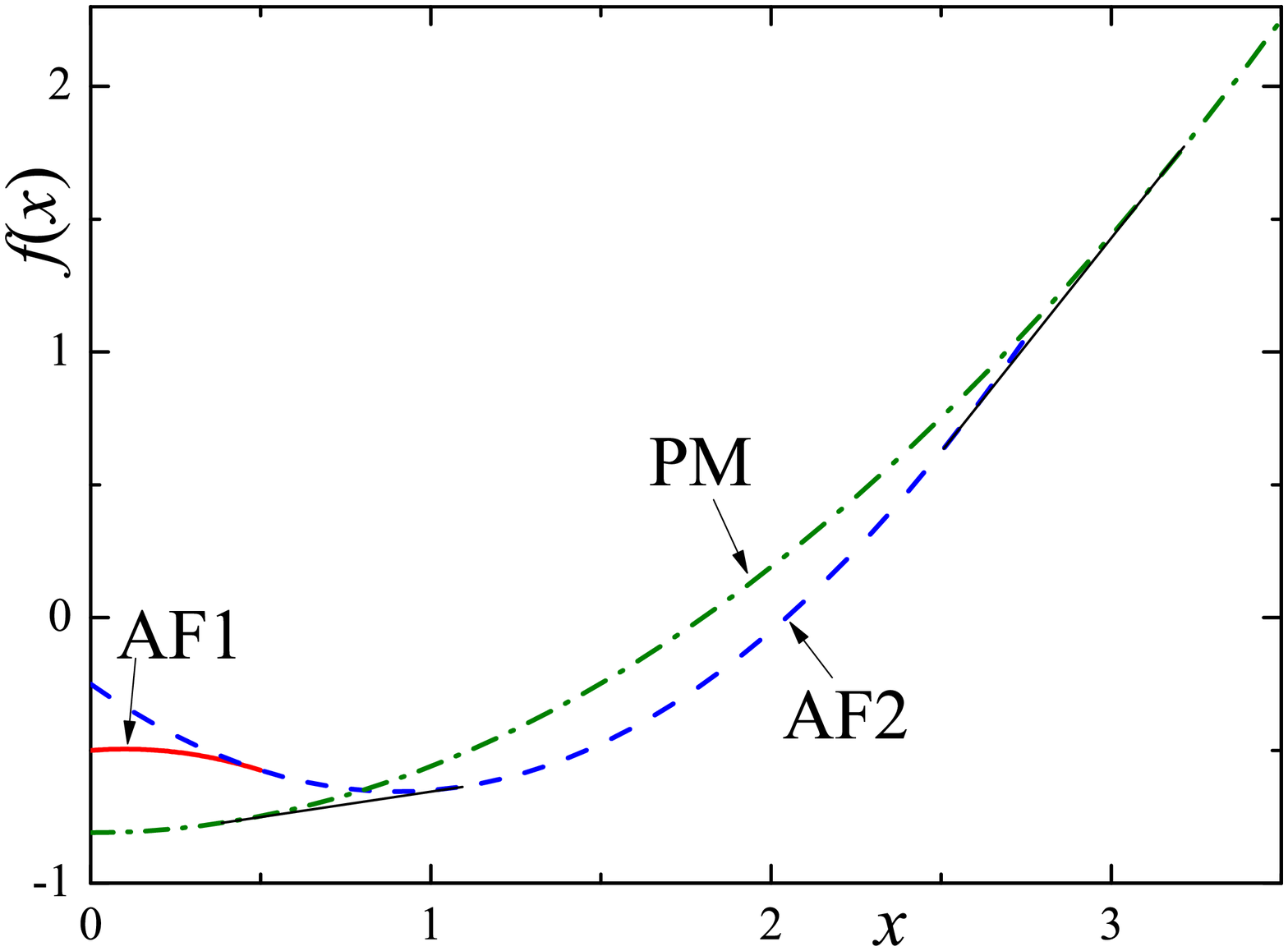}
\caption{(Color online) Dimensionless free energies $f$ for the phases AF1, AF2, and PM (for definition, see the text) versus doping $x$ calculated for $b=0.6$ (upper panel) and $b=1.2$ (lower panel). All other homogeneous phases have larger free energies at any doping level. Thin solid (black) lines show the free energies of the phase-separated states found by the Maxwell construction.
%%%%%%%%%%%%%%%%%%%%%%%%%%%%%%%%%%%%%%%%%%%
\label{FigFreeEnergy}
%%%%%%%%%%%%%%%%%%%%%%%%%%%%%%%%%%%%%%%%%%%%%%%%%%
}
\end{figure}

At zero temperature, we can replace the Fermi functions in the equations above by the Heaviside step function $f_F(\varepsilon)\rightarrow \Theta(-\varepsilon)$. After this substitution, the integrations in Eqs.~\eqref{dimenEq_LF_EHS} and~\eqref{GrPot_LB_EHS_dim} are easily performed and we obtain explicitly
\begin{eqnarray}
%%%%%%%%%%%%%%%%%%%%%%%%%%%%%%%%%%%%%%%%%%%
\label{T_0_main}
%%%%%%%%%%%%%%%%%%%%%%%%%%%%%%%%%%%%%%%%%%%%%%%%%%
\delta_\sigma&=&\sqrt{2|\nu_\sigma|-1}=\sqrt{1-2|x_\sigma|}\,,\nonumber\\
x_{\sigma}&=&-\frac{\partial \varphi_\sigma}{\partial \nu}
=\sgn(\nu_\sigma)(1-|\nu_\sigma|)\,,\quad x = \sum_\sigma x_\sigma,\nonumber\\
\varphi_{\sigma}&=&\frac14-|\nu_\sigma|+\frac{\nu_\sigma^2}{2}\,,
\end{eqnarray}
where $\nu_\sigma=\nu-\sigma b$ is a measure of the de-nesting in subsystem $\sigma$. Equations~(\ref{T_0_main}) are valid, when $|\nu_\sigma|>\delta_\sigma$ and $\delta_\sigma\neq0$. This state is metallic with a well-defined Fermi surface and we will refer to it as  AF$_\sigma^{\rm met}$.

When $|\nu_\sigma|<\delta_\sigma$, we derive from  Eqs.~\eqref{dimenEq_LF_EHS} and~\eqref{GrPot_LB_EHS_dim} that $\delta_\sigma=1$, $x_\sigma=0$, and $\varphi_\sigma=-1/4$. This is an insulating state with the gap in the electron spectrum. We will denote it as AF$_\sigma^{\rm ins}$.

In the paramagnetic state, $\delta_\sigma=0$ (further referred to as PM$_\sigma$), we obtain $x_\sigma=\nu_\sigma$ and  $\varphi_\sigma=-\nu_\sigma^2/2$.

The model is symmetric with respect to the sign of doping and the direction
of the magnetic field (up to the replacement
$\sigma\to-\sigma$).
Consequently, we can consider only the case of electron doping,
$x\geq0$,
and
$b\geq0$.

Thus, we have nine possible homogeneous phases:
(AF$_\uparrow^{\rm ins}$, AF$_\downarrow^{\rm ins}$),
(AF$_\uparrow^{\rm met}$, AF$_\downarrow^{\rm met}$),
(PM$_\uparrow$, PM$_\downarrow$),
(AF$_\sigma^{\rm ins}$, AF$_{-\sigma}^{\rm met}$),
(AF$_\sigma^{\rm ins}$, PM$_{-\sigma}$),
and (AF$_\sigma^{\rm met}$, PM$_{-\sigma}$),
where
$\sigma=\uparrow,\,\downarrow$.
We compared the free energies of these phases and found that only three of them can correspond to the ground states of the system ($b>0$):
AF1=(AF$_\uparrow^{\rm ins}$, AF$_{\downarrow}^{\rm met}$),
AF2=(AF$_\uparrow^{\rm ins}$, PM$_{\downarrow}$),
and PM=(PM$_\uparrow$, PM$_{\downarrow}$). The plots of free energies of these phases versus doping $x$ are shown in Fig.~\ref{FigFreeEnergy}. The phase diagram in the ($x,b$) plane for the homogeneous phases is shown in  the upper panel of Fig.~\ref{FigPhDiag}. Note that the phases AF1, AF2, and PM are metallic if $x \ne 0$: one subsystem ($\sigma = \downarrow$) for AF1 and AF2, and both subsystems for the PM phase have a Fermi surface.

\subsection{Phase separation}
%%%%%%%%%%%%%%%%%%%%%%%%%%%%%%%%%%%%%%%%%%%%%%%%%%
\label{subsect::phasep}
%%%%%%%%%%%%%%%%%%%%%%%%%%%%%%%%%%%%%%%%%%%%%%%%%%

The phase diagram discussed above was calculated neglecting the possibility
of phase separation. However, the shape of the $f(x)$ curves implies such a possibility near the transition lines between the homogeneous states [see solid (black) lines in Fig.~\ref{FigFreeEnergy}]. Indeed, the compressibility of the AF1 phase is negative, $\partial^2f/\partial x^2<0$ (see Fig.~\ref{FigFreeEnergy}), in the whole doping range where this phase exists. Thus, the homogeneous phase AF1 is unstable, and the separation into the AF1 phase with $x=0$ and AF2 phase occurs in the system. Let us refer to this phase-separated state as PS1. The range of doping where the phase-separated state is the ground state can be found using the Maxwell construction~\cite{Landau}. The analysis shows that the PS1 phase corresponds to the ground state of the system if $b<b_{c2}=1/\sqrt{2}$ and $0<x<1/\sqrt{2}$.

\begin{figure}[t]
\centering
\includegraphics[width=0.95\columnwidth]{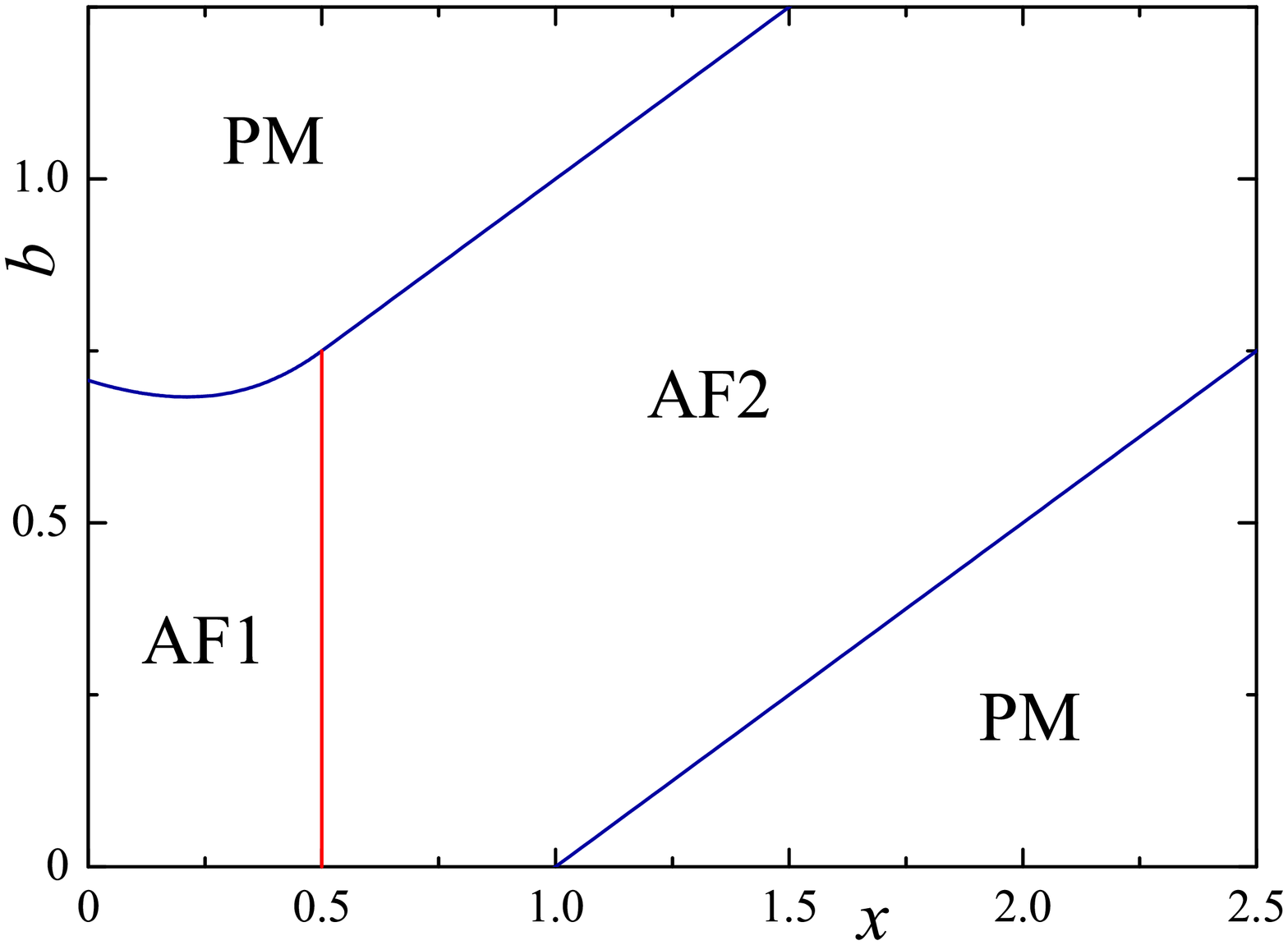}\\
\includegraphics[width=0.95\columnwidth]{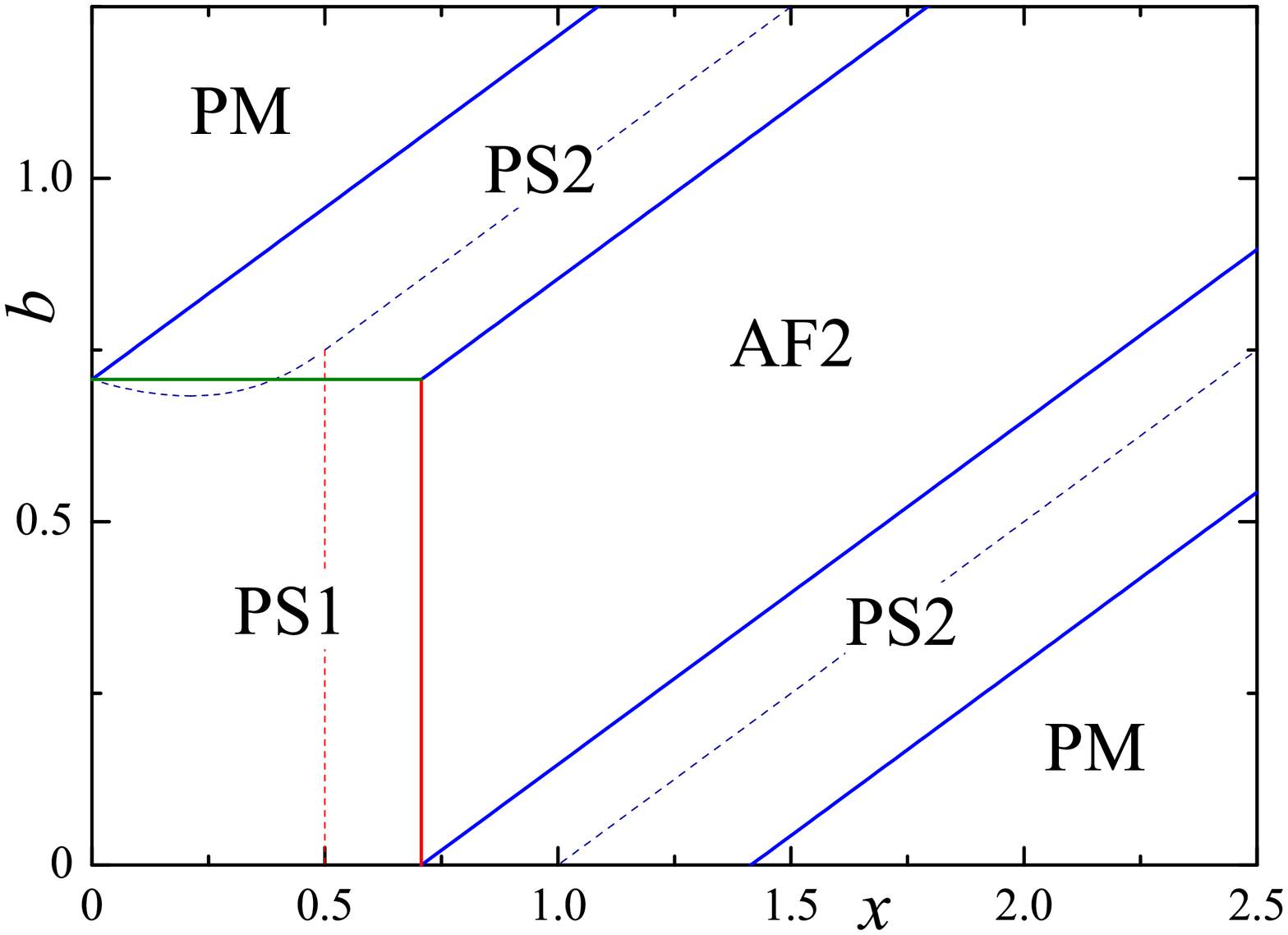}
\caption{(Color online) Phase diagram of the electron model at low fields and zero temperature, calculated neglecting the effect of the Landau quantization. In the upper panel, only homogeneous states are shown. The high-$b$ boundary of the AF1 phase is given by the equation $b = (x + \sqrt{4 x^2 - 4x + 2})/2$, while all other boundaries are straight lines. The phase diagram in the lower panel takes into account the possibility of phase separation. Thin dashed curves in the lower panel retrace the phase boundaries from the upper panel. The phase PS1 lies within the boundaries $b < 1/\sqrt{2}$ and $ x<1/\sqrt{2}$. At zero magnetic field the critical doping separating the PS2 and PM phases is equal to $x =\sqrt{2}$. The definitions of homogeneous phases are given in  Subsection~\ref{subsect::homogen}. The phase-separated states are defined in Subsection~\ref{subsect::phasep}.\label{FigPhDiag}}
\end{figure}

Other regions of the phase diagram, where an inhomogeneous phase is the ground state, appear in the vicinity of the line separating the AF2 and PM states. The corresponding inhomogeneous phase will be referred to as PS2. The phase PS2 corresponds to the ground state of the system within the doping range $2b+1/\sqrt{2}<x<2b+\sqrt{2}$, for any value of $b$, and also within the range $2b-\sqrt{2}<x<2b-1/\sqrt{2}$, if $b>b_{c2}$. The resulting phase diagram of the model is shown in the lower panel of Fig.~\ref{FigPhDiag}.

\section{Electron--hole coupling: high magnetic field}
%%%%%%%%%%%%%%%%%%%%%%%%%%%%%%%%%%%%%%%%%%%
\label{SectHighB}
%%%%%%%%%%%%%%%%%%%%%%%%%%%%%%%%%%%%%%%%%%%%%%%%%%

For higher magnetic field, we predict the existence of oscillations of the
SDW order parameter due to the Landau quantization. This phenomenon is
similar to the well known de Haas--van Alphen effect, manifesting itself in the oscillations of the magnetic moment in metals~\cite{Lifshitz}. To outline the general effects and to avoid excessive mathematical difficulties, we restrict our consideration to the case of ideal nesting at zero magnetic field and ideal electron--hole symmetry, that is, $x=0$ (the grand potential coincides with the free energy), $m_a=m_b=m$, $\Delta g=0$, $\varepsilon^a_{\text{max}}=\varepsilon^b_{\text{max}}\equiv\varepsilon_{\text{max}}$,
and
$\varepsilon^a_{\text{min}}=\varepsilon^b_{\text{min}}\equiv\varepsilon_{\text{min}}$.

Using Eqs.~\eqref{psi_expand} and~\eqref{hi_n}, we rewrite the interaction part of Hamiltonian \eqref{ham_int} in the form
\begin{eqnarray}
%%%%%%%%%%%%%%%%%%%%%%%%%%%%%%%%%%%%%%%%%%%
\label{H_int_1}
\hat{H}_{\textrm{int}}&=&\sum_{\sigma\sigma'}\sum_{nmn'm'}
\sum_{\mathbf{p}\mathbf{p}'\mathbf{q}}{V_{nmn'm'}
(\!p_x,p_x\!-\!q,p'_x,p'_x\!-\!q_x\!)}\times\nonumber\\
&&\psi^\dag_{\mathbf{p}na\sigma}\psi_{\mathbf{p}'n'a\sigma}
\psi^\dag_{\mathbf{p'-\mathbf{q}}m'b\sigma'}
\psi_{\mathbf{p-\mathbf{q}}mb\sigma'}\,,
\end{eqnarray}
where we introduce the matrix elements
\begin{eqnarray}
%%%%%%%%%%%%%%%%%%%%%%%%%%%%%%%%%%%%%%%%%%%
\label{vnmpq}
&&V_{nmn'm'}(\!p_x,p_x\!-\!q_x,p'_x,p'_x\!-\!q_x\!)=\frac{V}{{\cal V}^{2/3}l_B}\int\limits_{-\infty}^{+\infty}\!d\xi\times\nonumber\\
&&\chi_n(\xi-l_Bp_x)\chi_m[\xi-l_B(p_x-q_x)]\times\nonumber\\
&&\chi_{n'}(\xi-l_Bp'_x)\chi_{m'}[\xi-l_B(p'_x-q_x)]\,.
\end{eqnarray}
Let us remind that $\mathbf{p}$, $\mathbf{p}'$, and $\mathbf{q}$ in Eq.~\eqref{H_int_1}
are 2D momenta having $x$ and $z$ components.

In the mean-field approximation we apply the following replacement in the
interaction Hamiltonian
\begin{eqnarray}
%%%%%%%%%%%%%%%%%%%%%%%%%%%%%%%%%%%%%%%%%%%
\label{MFbreaking}
&&\psi^\dag_{\mathbf{p}na\sigma}
\psi^{\phantom{\dag}}_{\mathbf{p}'n'a\sigma}
\psi^\dag_{\mathbf{p'-\mathbf{q}}m'b\sigma'}
\psi^{\phantom{\dag}}_{\mathbf{p-\mathbf{q}}mb\sigma'}\to\\
&&\delta_{\mathbf{q0}}
\left[
	\eta_{nm\uparrow}(\mathbf{p})
	\eta^*_{n'm'\uparrow}(\mathbf{p'})
	+
	\eta_{nm\downarrow}(\mathbf{p})
	\eta^*_{n'm'\downarrow}(\mathbf{p'})
	-
	\phantom{\psi^\dag_{\mathbf{p'}m'b\uparrow}}
\right.
\nonumber
\\
&&\left(
	\eta_{nm\uparrow}(\mathbf{p})
	\psi^\dag_{\mathbf{p'}m'b\downarrow}
	\psi^{\phantom{\dag}}_{\mathbf{p'}n'a\uparrow}
	+
	\eta_{nm\downarrow}(\mathbf{p})
	\psi^\dag_{\mathbf{p'}m'b\uparrow}
	\psi^{\phantom{\dag}}_{\mathbf{p'}n'a\downarrow}
\right)
-
\nonumber
\\
&&\left.
\left(
	\eta^*_{n'm'\uparrow}(\mathbf{p'})
	\psi^\dag_{\mathbf{p}na\uparrow}
	\psi^{\phantom{\dag}}_{\mathbf{p}mb\downarrow}
	+
	\eta^*_{n'm'\downarrow}(\mathbf{p'})
	\psi^\dag_{\mathbf{p}na\downarrow}
	\psi^{\phantom{\dag}}_{\mathbf{p}mb\uparrow}
\right)
\right],
\nonumber
\end{eqnarray}
where we assume that mean values $\langle\psi^\dag_{\mathbf{p}na\sigma}
\psi^{\phantom{\dag}}_{\mathbf{p}'mb\bar{\sigma}}\rangle=0$ if
$\mathbf{p}\neq\mathbf{p}'$ and introduce the notation
\begin{equation}
%%%%%%%%%%%%%%%%%%%%%%%%%%%%%%%%%%%%%%%%%%%
\label{average_def}
\eta_{nm\sigma}(\mathbf{p})
=
\langle
	\psi^\dag_{\mathbf{p}na\sigma}
	\psi^{\phantom{\dag}}_{\mathbf{p}mb\bar{\sigma}}
\rangle
=
\langle
	\psi^\dag_{\mathbf{p}mb\bar{\sigma}}
	\psi^{\phantom{\dag}}_{\mathbf{p}na\sigma}
\rangle^*.
\end{equation}

Substitution \eqref{MFbreaking} makes the total Hamiltonian quadratic in the electron operators. As a result, we are able to calculate the electron spectrum and the grand potential of the system. Minimization of the grand potential with respect to $\eta_{nm\sigma}(\mathbf{p})$ would give us the infinite number of integral equations for the functions $\eta_{nm\sigma}(\mathbf{p})$. This procedure can be substantially simplified if we assume that the functions $\eta_{nm\sigma}(\mathbf{p})$ are independent of the momentum $p_x$. In other words, we assume here that the electron--electron interactions do not lift the degeneracy of the Landau levels, Eq.~\eqref{E_alpha}, with respect to the momentum $p_x$.
Making these assumptions, we effectively restrict the class of variational
mean-field wave functions, from which the approximate ground state wave
function is chosen. Without the latter simplifications, the calculations
become poorly tractable. Once this approximation is accepted, we obtain the following relation for the mean-field interaction Hamiltonian
\begin{equation}
%%%%%%%%%%%%%%%%%%%%%%%%%%%%%%%%%%%%%%%%%%%
\label{HamIntMf}
\hat{H}_{\textrm{int}}^{MF}\!\!=\!\!\sum_{p_x\sigma}\!\left[\!\frac{4\pi {\cal V}^{\scriptstyle\frac13}l_B^2\Delta_{\sigma}^2}{V}\!-\!
\sum_{p_zn}\!\left(\!\Delta_{\sigma}\psi^\dag_{\mathbf{p}nb\bar{\sigma}}
\psi^{\phantom{\dag}}_{\mathbf{p}na\sigma}\!+h.c.\right) \!\right],
\end{equation}
where the SDW order parameters $\Delta_{\sigma}$ now have the form
\begin{equation}
%%%%%%%%%%%%%%%%%%%%%%%%%%%%%%%%%%%%%%%%%%%
\label{Delta_HF-EHS}
\Delta_{\sigma}=\frac{V}{2\pi{\cal V}^{1/3}l_B^2}\sum_{p_zn}\eta_{nn\sigma}(p_z)\,.
\end{equation}
Thus, similar to the case of low magnetic fields considered in the previous
Section, we have two variational parameters to minimize the grand potential.

We diagonalize the total mean-field Hamiltonian
$\hat{H}_e+\hat{H}_{\textrm{int}}^{MF}$ and derive the expression for the grand potential of the system at zero doping (perfect nesting)
\begin{eqnarray}
%%%%%%%%%%%%%%%%%%%%%%%%%%%%%%%%%%%%%%%%%%%
\label{E00}
\Omega
&=&
{\cal V}^{1/3}\!
\sum_{p_x,\sigma}\!
	\left\{
		\frac{4\pi l_B^2\Delta_\sigma^2}{V}
		-
		\phantom{\frac{\sqrt{\Delta_\sigma^2}}{2}}
\right.
\\
&&
\left.
		2T\sum_{n}\int\!\!\frac{dp_z}{2\pi}
				\ln\!\!\left[
					2\cosh\!\left(
						\frac{
						\sqrt{
							\Delta_\sigma^2\!
							+\!
							\varepsilon_\sigma^2
							\left(p_z,n\right)
						}}{2T}
						\right)
					\right]
	\right\},
\nonumber
\end{eqnarray}
where
%-\!\sum_{n}\int\!\!\frac{dp_z}{2\pi} \sqrt{\Delta_\sigma^2\!+\!\varepsilon_\sigma^2\left(p_z,n\right)}\right],
\begin{equation}
%%%%%%%%%%%%%%%%%%%%%%%%%%%%%%%%%%%%%%%%%%%%%%%%%%
\label{epsilon}
%%%%%%%%%%%%%%%%%%%%%%%%%%%%%%%%%%%%%%%%%%%%%%%%%%
\varepsilon_\sigma(p_z,n)
=
\omega_H\!\!\left(n+\frac{1}{2}\right)+\frac{p_z^2}{2m}-E_{F\sigma}\,.
\end{equation}
In Eq.~\eqref{E00}, the summation over $n$ and the integration over
$p_z$ are taken within the range determined by the inequalities $\varepsilon_{\text{min}} < \varepsilon_{\sigma}(p_z,n) <\varepsilon_{\text{max}}$.

The summation over $n$ in Eq.~\eqref{E00} can be replaced by the integration over the 2D momentum $\mathbf{p}=(p_x,\,p_y)$, when the distance between Landau levels is smaller than the SDW band gap ($\omega_H\ll\Delta_0$). In this case we have
\begin{equation*}
n\to\frac{\mathbf{p}^2l_B^2}{2}\,,\;\;\frac{1}{l_B^2}
\sum_n\ldots\to\int\!d\!\left(\!\frac{p^2}{2}\!\right)\ldots
=
\int\frac{dp_xdp_y}{2\pi}\ldots
\end{equation*}
As a result, Eq.~\eqref{E00} is replaced by Eq.~\eqref{GrPot}, where the integration is performed over 3D momentum. This justifies the assumption made in the previous Section that we can neglect the effect of the Landau level quantization at low fields.

Minimization of the potential $\Omega$ gives the equation for the gap
\begin{equation}
%%%%%%%%%%%%%%%%%%%%%%%%%%%%%%%%%%%%%%%%%%%
\label{Gap00}
\frac{1}{4\pi^2l_B^2}\sum_{n}\int\!dp_z\frac{\tanh\!
\left(\!\sqrt{\Delta_\sigma^2\!+\!\varepsilon_\sigma^2
\left(p_z,n\right)}/2T\right)}
{\sqrt{\Delta_\sigma^2\!+\!\varepsilon_\sigma^2\left(p_z,n\right)}}=\frac{2}{V}\,.
\end{equation}

We introduce the density of states
\begin{equation}
%%%%%%%%%%%%%%%%%%%%%%%%%%%%%%%%%%%%%%%%%%%%%%%%%% 
\label{DOS_def}
%%%%%%%%%%%%%%%%%%%%%%%%%%%%%%%%%%%%%%%%%%%%%%%%%%
\rho_{B}(E)\!=\!\frac{1}{4\pi^2l_B^2}\sum_n\!\int\!\! dp_z\delta\!\left[E-\omega_H\!\!\left(\!n\!+\!\frac{1}{2}\!\right)-\frac{p_z^2}{2m}\right],
\end{equation}
and rewrite Eq.~\eqref{Gap00}
in the form
\begin{equation}
%%%%%%%%%%%%%%%%%%%%%%%%%%%%%%%%%%%%%%%%%%%%%%%%%% 
\label{Gap0DOS}
%%%%%%%%%%%%%%%%%%%%%%%%%%%%%%%%%%%%%%%%%%%%%%%%%% 
\!\!\!\int\limits_{-E_{F\sigma}}^{\varepsilon_{\text{max}}
-E_{F\sigma}}\!\!\!\!\!\!\!\!\!\!d\varepsilon\,\rho_{B}
(\varepsilon+E_{F\sigma})\frac{\tanh\!\left(\!\sqrt{\Delta_\sigma^2\!+
\!\varepsilon^2}/2T\right)}
{\sqrt{\Delta_\sigma^2+\varepsilon^2}}=\frac{2}{V}\,.
\end{equation}

The density of states exhibits equidistant peaks at energies $E=\omega_H(n+1/2)$. This results in the oscillatory behavior of the order parameters $\Delta_\sigma$ on the magnetic field similar to the de Haas--van Alphen effect~\cite{Lifshitz}. In the limit $\omega_H/\varepsilon_{F}\ll1$,
one can calculate the density of states analytically. Details of the
calculations are presented in the Appendix, where for $\rho_{B}(E)$ we derive expression \eqref{DOS_fourier}. Substituting this expression into
Eq.~\eqref{Gap0DOS} at $T=0$, we obtain
\begin{eqnarray}
%%%%%%%%%%%%%%%%%%%%%%%%%%%%%%%%%%%%%%%%%%%
\label{GAP_FIN}
%%%%%%%%%%%%%%%%%%%%%%%%%%%%%%%%%%%%%%%%%%%%%%%%%%
&&\ln\!\left(\!\frac{\Delta_{\sigma}}{\Delta_0}\!\right)
+\sqrt{\frac{\omega_H}{2\varepsilon_F}}\times\\
\nonumber
&&\sum_{l=1}^{\infty}\frac{(-1)^l}{\sqrt{l}}\cos\!\left(\!\frac{2\pi lE_{F\sigma}}{\omega_H}-\frac{\pi}{4}\right)K_0\!\left(\!\frac{2\pi l \Delta_{\sigma}}{\omega_H}\!\right)=0\,,
\end{eqnarray}
where
$K_0(z)$
is the Macdonald function of zeroth order. We solve this equation
numerically. Since the Macdonald functions decay exponentially at large
values of their arguments, the series in Eq.~\eqref{GAP_FIN} converges quickly.

\begin{figure}
\begin{center}
\includegraphics[width=0.95\columnwidth]{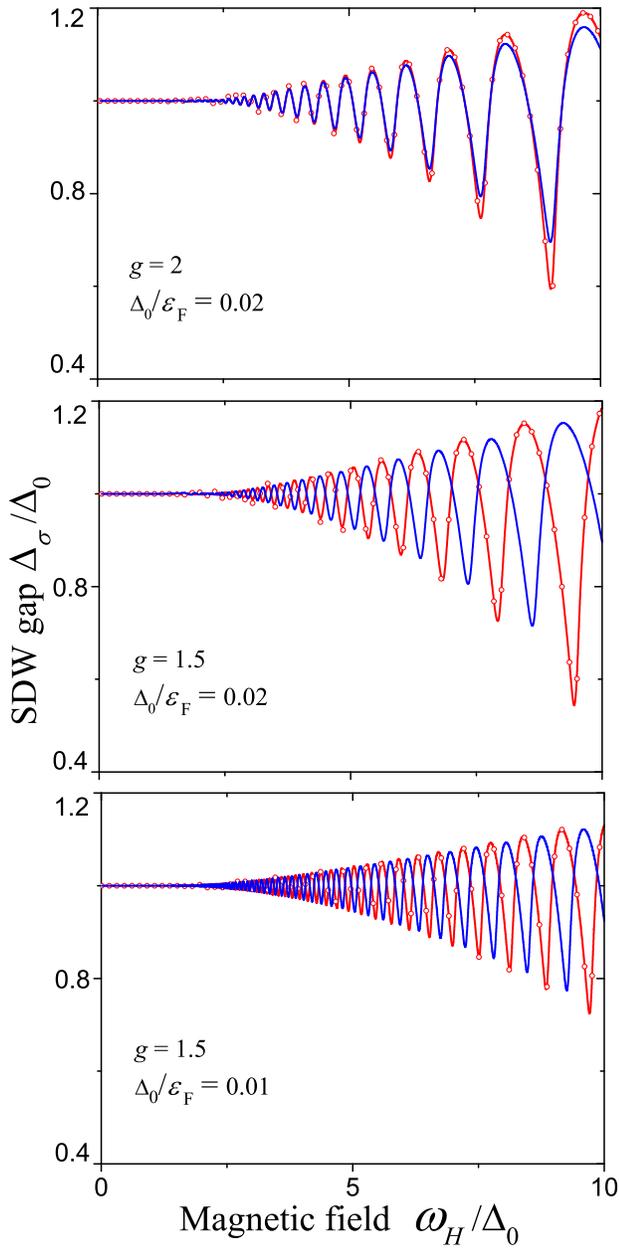}
\end{center}
\caption{(Color online) SDW gaps versus magnetic field calculated at different ratios $\Delta_0/\varepsilon_{F}$ and different values of $g$ specified in the plots; $\Delta_\uparrow$ is shown by the (red) solid line with open circles, while $\Delta_\downarrow$ by the (blue) solid line without symbols.
%%%%%%%%%%%%%%%%%%%%%%%%%%%%%%%%%%%%%%%%%%%%%%%%%%
\label{Gap}
%%%%%%%%%%%%%%%%%%%%%%%%%%%%%%%%%%%%%%%%%%%%%%%%%%
}
\end{figure}

The calculated parameters $\Delta_\sigma(B)$ for different $g$ are shown in Fig.~\ref{Gap}. We see that both order parameters $\Delta_\sigma(B)$
oscillate when the magnetic field is varied. The amplitudes of the
oscillations increase when the ratio $\Delta_0/\varepsilon_F$ grows (that
is, the interaction increases). The oscillations of $\Delta_\uparrow$ and
$\Delta_\downarrow$ have the same phases, if $g$ is an integer, and
different phases otherwise.

It is seen in Fig.~\ref{Gap} that the order parameters oscillate about some mean value $\widetilde \Delta_\sigma$, which is quite robust against the growth of $B$. This stability, however, is a consequence of the perfect nesting. The value of $\widetilde \Delta_\sigma$ decreases with $B$ if we take into account either doping or electron--hole asymmetry. When the doping or asymmetry is high, the SDW order disappears before pronounced oscillations arise.

Note also that the SDW phase is stable at low temperatures since when $T\rightarrow0$ the free energy of the magnetically-ordered phase is lower than the PM one. This can be checked directly.

\begin{figure}
\begin{center}
\includegraphics[width=1\columnwidth]{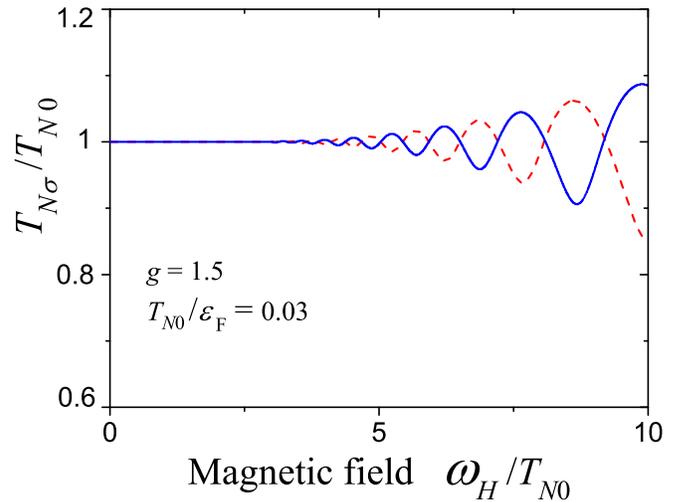}
\end{center}
\caption{(Color online) Dependence of the N\'eel temperatures on the magnetic field; $T_{N\uparrow}$ is shown by (red) dashed line, while $T_{N\downarrow}$
by (blue) solid one. Parameters are specified in the plot.
%%%%%%%%%%%%%%%%%%%%%%%%%%%%%%%%%%%%%%%%%%%
\label{figTN}
%%%%%%%%%%%%%%%%%%%%%%%%%%%%%%%%%%%%%%%%%%%%%%%%%%
}
\end{figure}

In addition to the behavior of the order parameters, the temperatures of
the phase transitions also oscillate as a function of the magnetic field.
Note that for $B\neq0$, there are two transition temperatures,
$T_{N\sigma}$, where, as usual, $\sigma = \uparrow, \downarrow$. These
temperatures can be calculated using
Eq.~\eqref{Gap0DOS}
by taking the limit
$\Delta_\sigma\to0$.
This gives the equation
\begin{equation}
%%%%%%%%%%%%%%%%%%%%%%%%%%%%%%%%%%%%%%%%%%%%%%%%%% 
\label{TN0DOS}
%%%%%%%%%%%%%%%%%%%%%%%%%%%%%%%%%%%%%%%%%%%%%%%%%% 
\int\!d\varepsilon\,\rho_{B}(\varepsilon+E_{F\sigma})\frac{\tanh\!
\left(\!\varepsilon/2T_{N\sigma}\right)}{\varepsilon}=\frac{2}{V}\,.
\end{equation}
Using the density of states \eqref{DOS_fourier}, we obtain
\begin{eqnarray}
\nonumber\label{TNfin}
\ln\frac{T_{N\sigma}}{T_{N0}}&=&\sqrt{\frac{\omega_H}{2\varepsilon_F}}
\sum\limits_{l=1}^{\infty}\frac{(-1)^l}{\sqrt{l}}\cos\!\left(\!\frac{2\pi lE_{F\sigma}}{\omega_H}-\frac{\pi}{4}\!\right)\times\\
&&\ln{\!\left[\tanh\left(\frac{\pi^2lT_{N\sigma}}{\omega_H}\right)\right]}\,,
\end{eqnarray}
where $T_{N0}$ is the N\'{e}el temperature at zero field, which is related to the SDW gap according to the BCS-like formula $T_{N0}\cong0.567\Delta_0$. The results of these calculations are shown in Fig.~\ref{figTN}.

\section{Discussion}
%%%%%%%%%%%%%%%%%%%%%%%%%%%%%%%%%%%%%%%%%%%
\label{disco}
%%%%%%%%%%%%%%%%%%%%%%%%%%%%%%%%%%%%%%%%%%%%%%%%%%

In this work, we investigated the effect of an applied magnetic field on weakly-correlated electron systems with imperfect nesting. Such study may be relevant for recent experiments on doped rare-earth borides~\cite{Sluchanko_PRB2015}. We found that, when the cyclotron frequency $\omega_\alpha$ is comparable to the electron energy gap, $\Delta_0$, the magnetic field effects must be taken into account.

The magnetic field enters the model Hamiltonian via two channels: (i) the
Zeeman term, and (ii) orbital (or, diamagnetic) contribution. At low field, $\omega_\alpha<\Delta_0$, and not too small Land\'e factors $g_\alpha$, one
can neglect the latter contribution and take into account only the Zeeman term. We investigated the combined effects of both terms in the limit of ideal electron--hole symmetry and ideal nesting.

Our study demonstrated that in the presence of the Zeeman term, the number
of possible homogeneous magnetically-ordered phases significantly
increases, compared to the case of
$B=0$.
In
Subsection~\ref{subsect::homogen},
we defined as many as nine possible states with different symmetries. If
necessary, this list may be increased by taking into account incommensurate
SDW
phases~\cite{Rice}
and phases with
``stripes''~\cite{zaanen_stripes1989,tokatly1992}.
Of this abundance, only two ordered homogeneous phases could serve as a
ground state of our model.

When inhomogeneous states are included into consideration, even the
zero-temperature phase diagram becomes quite complex. We would like to
remind a reader that, theoretically, the phase separation is a very robust
phenomenon. Its generality goes beyond the weak-coupling nesting
instabilities of a Fermi surface: the phase separation is found in
multi-band Hubbard and Hubbard-like  models, where the nesting is not
crucial~\cite{bascones2012,dagotto2014,WePRL2005,WePRB2007,We_grA_grE2012}.
It is therefore important to account for its possibility both
theoretically, and experimentally.

However, phase separation is not universal: the simplifications of our
approach, and the contributions, which we have intentionally omitted
(Coulomb interaction, lattice effects, realistic shape of the Fermi
surface, disorder, etc), can restore the stability of
homogeneous states for a given set of model parameters. For example, the long-range Coulomb repulsion, caused by the charge redistribution in inhomogeneous state, suppresses the phase separation~\cite{DiCastro1,DiCastro2,BianconiArrested}.
Therefore, on experiments the inhomogeneous states may occupy fairly modest
part of the phase diagram, as seen, for example, in
Refs.~\onlinecite{phasep_exp2012,Narayanan_RRL2014}.
The final location of the segregated region on the phase diagram is
affected by the Zeeman energy, as our calculations demonstrated.

The orbital contribution to the Hamiltonian leads to the Landau quantization of
the single-particle orbits. As a result, we have demonstrated that both
order parameters and the N\'eel temperatures oscillate as the magnetic
field changes. This behavior is associated with the oscillatory part of the
single-particle density of states, which emerges due to the Landau
quantization. The same oscillations of the density of states are also the
cause of the de Haas--van Alphen effect. Yet another related phenomenon,
the so-called field-induced SDW, is known to occur in quasi-one-dimensional materials~\cite{GorkovLebed_JdePhL1984_magn_field_stab1D,
Lebed_JETP1985_magn_field_PhDiag,Yakovenko_JETP1987_field_ind_PhTr,
Azbel_PRB1989_field_ind_SDW,Pudalov_Springer2008_magn_field_SDW,
Lebed_Springer2008_magn_field_SDW}.

Pronounced oscillations of both $\Delta_\sigma$ and $T_{N \sigma}$
develop at sufficiently large magnetic fields. This circumstance makes  their experimental observation a delicate issue. Indeed, the results of Section~\ref{SectHighB} were obtained under the assumption of perfect electron--hole symmetry. In a more realistic case, this symmetry is broken, and the magnetic field may cause a transition into a phase with a different order parameter, or destroy the SDW completely before an observable oscillatory trend sets in.

We demonstrated that in electron systems with imperfect nesting the applied magnetic field leads to a significant increase in the number of possible ordered states. It also affects the inhomogeneous, phase-separated states of the system. At higher fields, the Landau quantization causes oscillations of the SDW order parameters and of the corresponding N\'eel temperatures.

\section*{Acknowledgements}
%%%%%%%%%%%%%%%%%%%%%%%%%%%%%%%%%%%%%%%%%%%
\label{ackno}
This work is partially supported by the Russian Foundation for Basic
Research (projects 14-02-00276 and 15-02-02128), Russian Ministry of
Education and Science (grant 14.613.21.0019 (RFMEFI61314X0019)),
RIKEN iTHES Project,the MURI Center for Dynamic Magneto-Optics via the
AFOSR award number FA9550-14-1-0040, the IMPACT program of JST, a
Grant-in-Aid for Scientific Research (A), and a grant from the John
Templeton Foundation.

\appendix*

\section{Density of states and the equation for the gap}

Here, we derive the explicit expression for the density of states given by
Eq.~\eqref{DOS_def}. The main idea is to divide the density of states into  monotonic and oscillatory parts (a similar approach is used, e.g., in Chapter~6 of Ref.~\onlinecite{Lifshitz}). To do this, we first take the integral over $p_z$ in Eq.~\eqref{DOS_def}, which gives
\begin{equation}
%%%%%%%%%%%%%%%%%%%%%%%%%%%%%%%%%%%%%%%%%%%
\label{DOS_1}
%%%%%%%%%%%%%%%%%%%%%%%%%%%%%%%%%%%%%%%%%%%%%%%%%%
\rho_B(E)
=
\frac{1}{4\pi^2l_B^2}\sqrt{\frac{2m}{\omega_H}}
\sum_{n=0}^{N_0}{\frac{1}{\sqrt{n+a}}},
\end{equation}
where
\begin{equation}
%%%%%%%%%%%%%%%%%%%%%%%%%%%%%%%%%%%%%%%%%%%
\label{N_a}
%%%%%%%%%%%%%%%%%%%%%%%%%%%%%%%%%%%%%%%%%%%%%%%%%%
N_0
=
\left[\frac{E}{\omega_H}-\frac{1}{2}\right],
\quad
a=\frac{E}{\omega_H}-\frac{1}{2}-N_0\,,
\end{equation}
and $[\dots]$ denotes the integer part of a real number. The parameter $a$,
by its definition, is the fractional part of $E/\omega_H - 1/2$. Using
formulas from the theory of Euler integrals,
\begin{equation}\nonumber
\frac{1}{\sqrt{x}}\!=\!\frac{1}{\sqrt{\pi}}\!\int\limits_0^\infty\!
\frac{ds}{\sqrt{s}}e^{-xs},\,\,\, 2\sqrt{x}\!=\!\frac{1}{\sqrt{\pi}}\!\int\limits_0^\infty \!\frac{ds}{s^{3/2}}\left(1-e^{-xs}\right)\,,
\end{equation}
we write the sum in
Eq.~\eqref{DOS_1}
in the form
\begin{equation}
\sum_{n=0}^{N_0}{\frac{1}{\sqrt{n+a}}}=2\sqrt{N_0+a}+F(N_0+a)+g(a)\,,
\end{equation}
where
\begin{eqnarray}
%%%%%%%%%%%%%%%%%%%%%%%%%%%%%%%%%%%%%%%%%%%
\label{sum}
%%%%%%%%%%%%%%%%%%%%%%%%%%%%%%%%%%%%%%%%%%%%%%%%%%
F(N_0+a)\!&=&\!\frac{1}{\sqrt{\pi}}\!\int\limits_0^\infty \!\!\frac{ds}{\sqrt{s}}\left(\frac{1}{s}\!+\!\frac{1}{1-e^{s}}
\right)e^{-(N_0+a)s},\nonumber\\
g(a)\!&=&\!\frac{1}{\sqrt{\pi}}\!\int\limits_0^\infty \!\!\frac{ds}{\sqrt{s}}\left(\frac{e^{-as}}{1-e^{-s}}\!-\!\frac{1}{s}\right)\,.
\end{eqnarray}

In the limit
$\omega_H/\varepsilon_F\ll 1$
one has
\begin{equation*}
F(N_0+a)\approx \frac{1}{2\sqrt{N_0+a}}+O\left(\frac{1}{N_0^{3/2}}\right).
\end{equation*}
Substituting all these expressions into Eq.~\eqref{DOS_1} and expanding it in powers of $\omega_H/\varepsilon_F\ll 1$, we obtain
\begin{equation}
%%%%%%%%%%%%%%%%%%%%%%%%%%%%%%%%%%%%%%%%%%%
\label{DOS_decay}
%%%%%%%%%%%%%%%%%%%%%%%%%%%%%%%%%%%%%%%%%%%%%%%%%%
\rho_{B}(E)\cong\rho_0(E)+\frac{(2m)^{3/2}\sqrt{\omega_H}}{8\pi^2}g[a(E)]+
O\!\left(\frac{\omega^2_H}{\varepsilon^2_{F\sigma}}\right)\,,
\end{equation}
where
\begin{equation}
%%%%%%%%%%%%%%%%%%%%%%%%%%%%%%%%%%%%%%%%%%%
\label{DOS_0}
%%%%%%%%%%%%%%%%%%%%%%%%%%%%%%%%%%%%%%%%%%%%%%%%%%
\rho_0(E)=\frac{(2m)^{3/2}\sqrt{E}}{4\pi^2}
\end{equation}
is the density of states at zero magnetic field.

The second term in the right-hand side of
Eq.~(\ref{DOS_decay})
oscillates as $B$ changes, since $a$ is an oscillating function of $B$, see
Eq.~(\ref{N_a}).
We expand the function
$g(a)$
in a Fourier series
\begin{equation}
%%%%%%%%%%%%%%%%%%%%%%%%%%%%%%%%%%%%%%%%%%%
\label{g_fur}
%%%%%%%%%%%%%%%%%%%%%%%%%%%%%%%%%%%%%%%%%%%%%%%%%%
g(a)=\!\!\!\sum\limits_{l=-\infty}^{+\infty}g_le^{2i\pi la},\quad g_l=\int_0^1e^{-2i\pi la}g(a)\,,
\end{equation}
where the coefficients of the series can be calculated analytically
\begin{equation}
%%%%%%%%%%%%%%%%%%%%%%%%%%%%%%%%%%%%%%%%%%%
\label{g_n}
%%%%%%%%%%%%%%%%%%%%%%%%%%%%%%%%%%%%%%%%%%%%%%%%%%
g_0=0,\,\;\;g_l=\frac{1-i\textrm{sgn}(l)}{2}\frac{1}{\sqrt{|l|}}\;(l\neq0)\,.
\end{equation}
Keeping the leading corrections, we can write
\begin{eqnarray}
%%%%%%%%%%%%%%%%%%%%%%%%%%%%%%%%%%%%%%%%%%%
\label{DOS_fourier}
%%%%%%%%%%%%%%%%%%%%%%%%%%%%%%%%%%%%%%%%%%%%%%%%%%
\rho_{B}(E)
&\cong&
\rho_0(E)
+
\frac{\rho_0(\varepsilon_F)}{4}\sqrt{\frac{\omega_H}{\varepsilon_F}}
\times\\
&&
\sideset{}{'}\sum\limits_{l=-\infty}^{\infty}
	\frac{1\!-\!i\textrm{sgn}(l)}{\sqrt{|l|}}(-1)^l\exp\!\!{\left(\!\frac{2i\pi lE}{\omega_H}\!\right)}\,,
\nonumber
\end{eqnarray}
where the prime at the summation sign implies that the term with
$l=0$ has to be omitted. Now, we substitute
Eqs.~\eqref{DOS_decay}\,--\,\eqref{g_n}
in
Eq.~\eqref{Gap00}
and obtain the equation for the order parameter in the form
\begin{eqnarray}
%%%%%%%%%%%%%%%%%%%%%%%%%%%%%%%%%%%%%%%%%%%
\label{BCS_prom}
%%%%%%%%%%%%%%%%%%%%%%%%%%%%%%%%%%%%%%%%%%%%%%%%%%
&&\frac{2}{V}=\!\!\!\!\!\!\!
\int\limits_{-E_{F\sigma}}^{\varepsilon_{\text{max}}-E_{F\sigma}}
\!\!\!\!\!\!\!\!
\frac{d\varepsilon}{\sqrt{\Delta_\sigma^2+\varepsilon^2}}
\left\{
	\rho_0(\varepsilon+E_{F\sigma})
	+
	\frac{\rho_0(\varepsilon_F)}{4}\sqrt{\frac{\omega_H}{\varepsilon_F}}
	\times\phantom{\!\!\!\!\!\!\!\!\!\!\sum\limits_1^2}
\right.
\nonumber
\\
&&\left.
	\sideset{}{'}
	\sum\limits_{l=-\infty}^{\infty}
	\frac{1\!-\!i\textrm{sgn}(l)}{\sqrt{|l|}}(-1)^l
	\exp\!{
		\left(\!
			\frac{2i\pi l(\varepsilon+E_{F\sigma})}{\omega_H}\!
		\right)}
	\right\}.
\end{eqnarray}

As a consistency check, let us consider the limit of vanishing magnetic
field. In this case only the first term in the integral above survives, and
one obtains
\begin{equation}
\int\limits_{-E_{F\sigma}}^{\varepsilon_{\text{max}}-E_{F\sigma}}
\!\!\!\!\!\!\!\!\!\!
d\varepsilon\,\frac{\rho_0(\varepsilon+E_{F\sigma})}
{\sqrt{\Delta_\sigma^2+\varepsilon^2}}
\cong2\rho_0(\varepsilon_F)\ln\frac{2\sqrt{\varepsilon_F
(\varepsilon_{\text{max}}-\varepsilon_F)}}{\Delta_\sigma}\,.
\end{equation}
Thus, at zero magnetic field we have
\begin{equation}
\frac{1}{V}=\rho_0(\varepsilon_F)\ln\frac{2\sqrt{\varepsilon_F
(\varepsilon_{\text{max}}-\varepsilon_F)}}{\Delta_0}\,.
\end{equation}
Assuming that the Fermi level lies near the center of the electron or hole bands ($\varepsilon_F\sim\varepsilon_{\text{max}}/2$), we reproduce Eq.~\eqref{BCS} for the $\Delta_0$.

The integration of the second term in Eq.~\eqref{BCS_prom}
can be extended over all real values of $\varepsilon$ from $-\infty$ to $+\infty$. Taking this integral, we arrive finally to the
formula~\eqref{GAP_FIN} for the gap equation. Equations~\eqref{TNfin} for the N\'{e}el temperatures can be obtained in a similar manner.

\end{document}